\documentstyle[aps,epsf,twocolumn]{revtex}
\begin{document}
\title{On continuous-variable entanglement with and without phase references}

\author{S.J. van Enk and Terry Rudolph\\
Bell Labs, Lucent Technologies\\
600-700 Mountain Ave\\
Murray Hill NJ 07974}
\maketitle
\begin{abstract}
We discuss how continous-variable quantum states such as coherent states and two-mode squeezed states can be encoded in phase-reference independent ways.
\end{abstract}
\medskip

 By encoding quantum information in quantum states invariant under transformations between different reference frames (defining the three 
spatial directions), one can perform quantum communication protocols without having to share reference frames, as was pointed out in Ref.~\cite{bartlett}.

For continuous-variable quantum communication (and teleportation \cite{akira} in particular) it is not so much a reference frame but a phase reference (or equivalently, synchronized clocks) that is needed. This need arises from the fact that one tries to encode information in superpositions of different energy eigenstates \cite{jmo}. In particular, for light beams one would like to create superpositions of different numbers of photons.
Inspired by Ref.~\cite{bartlett} we consider here encoding schemes that apparently eliminate the need for a shared phase reference. 

The basics of ``phase-reference-free'' encoding is trivial: instead of using number states $|n\rangle$ one uses states of the form
\begin{equation}\label{code}
|n_M\rangle\equiv |n\rangle |M-n\rangle,
\end{equation}
where the ``logical'' state $|n_M\rangle$ is encoded in two modes of the same frequency, containing a fixed number $M$ of photons. More generally, we could encode in states with fixed total energy. 
With this type of encoding the basis states contain no time-dependent phase factors.

Now all one has to do, in principle, to perform a continuous-variable quantum protocol without using a phase reference, is to replace every relevant mode by two modes and encode according to (\ref{code}).
The only assumption here is that two parties who do not share a phase standard, will agree on the definition of number states.

What we do here is consider one particular way of achieving the encoding (\ref{code}), starting from standard quantum-optical states. 
The procedure used will explicitly demonstrate how synchronized clocks are {\em apparently} no longer needed, for example, to establish an entangled state.

The two types of states featuring in the Caltech teleportation experiment \cite{akira} are coherent states and the two-mode squeezed state.
Suppose we have a perfect phase reference available with a phase $\phi$ and frequency $\omega$ to which all our states are phase-locked. 
Then we would write down the state of a laser field as 
a coherent state
\begin{equation}\label{coherent}
|\alpha_{\phi}\rangle=\exp(-|\alpha|^2/2)
\sum_n \frac{\alpha^n e^{in\phi}}{\sqrt{n!}}|n\rangle,
\end{equation}
where $\alpha$ may still contain a phase, which is just the phase relative to the phase reference. The subscript $\phi$ reminds us that we can only write down a coherent state if we have the phase reference available \cite{laser}.

Now in order to create a two-mode squeezed state, we would first apply a frequency doubling. This is effected by a Hamiltonian of the form $H\propto a^2 b^{\dagger}+H.c.$, with $a$ describing a mode of 
frequency $\omega$ and $b$ a mode of frequency $2\omega$. 
In a coherent state of mode $a$ we may replace $a$ by the coherent state amplitude $\alpha\exp(i\phi)$. The result is
that a frequency doubled  coherent state 
contains a phase $2\phi$. 
Using such a state then to produce a two-mode squeezed state through a Hamiltonian of the form $H\propto b a_1^{\dagger}a_2^{\dagger}$
(with $a_{1,2}$ describing modes of frequency $\omega$) would give rise to a state of the form
\begin{equation}\label{two}
|\eta_{\phi}\rangle=\sqrt{1-\eta^2}
\sum_n \eta^n\exp(2in\phi) |n\rangle|n\rangle.
\end{equation}
The subscript $\phi$ reminds us once again that we only may write down this state if we have a phase reference available.

Now, however, consider how we can produce states whose form is independent of the value of $\phi$, and hence independent of the presence of a phase reference.
First, here is how one might achieve the encoding (\ref{code}) for
a coherent state $|\alpha_{\phi}\rangle$. 
Take an ancilla mode of the same frequency $\omega$ prepared in a coherent state
$|\beta_{\phi}\rangle$, 
and perform a QND measurement of the total photon number in both modes together. Of course, this is not something that is easy to do at all in practice, but here we are only concerned with a matter of principle.
Suppose the outcome is $M$. Then the joint state reduces to (in terms of encoded states)
\begin{eqnarray}\label{ab}
|\Psi\rangle&=&
\frac{1}{\sqrt{{\cal N}}}\sum_n \frac{(\alpha/\beta)^n}{\sqrt{n!(M-n)!}}
|n_M\rangle,\nonumber\\
{\cal N}&=&\sum_n \frac{|\alpha/\beta|^{2n}}{n!(M-n)!}.
\end{eqnarray}
Obviously, the state $|\Psi\rangle$ is no longer a coherent state in terms of the logical states $|n_M\rangle$. On the other hand, it is clearly independent of the phase $\phi$. Moreover, when $|\beta|$ is much larger than $|\alpha|$, the state $|\Psi\rangle$ does approximate a coherent state. Intuitively, this is because in that limit the measurement of the total number of photons hardly gives us any information about the original coherent state, as almost all photons originate from the second coherent state. Mathematically, we may approximate $(M-n)!$ by $M!/M^n$ in Eq.~(\ref{ab}) (for small values of $n$) in that limit to obtain a coherent state with amplitude $\alpha'=\alpha 
\sqrt{M}/\beta$. As $M$ will have a value around $|\beta|^2$, $|\alpha'|\approx|\alpha|$. Thus, for large $\beta$ we get
\begin{eqnarray}\label{ab2}
|\Psi\rangle\approx 
\exp(-|\alpha'|^2/2)\sum_n \frac{(\alpha')^n}{\sqrt{n!}}
|n_M\rangle.
\end{eqnarray}
Next, consider how we encode the two-mode squeezed state $|\eta_{\phi}\rangle$.
Add two ancilla modes in two
identical coherent states $|\beta_{\phi}\rangle$ with frequency $\omega$, and
perform two QND measurements of total photon number, each on two modes. In each case,  one mode contains a coherent state, the other is half of the two-mode squeezed state. Assume that the outcomes are $K$ and $L$,  respectively.
The state then reduces to, in terms of the encoded states (\ref{code}),
\begin{eqnarray}\label{ML}
|\Phi_{KL}\rangle&=&\frac{1}{\sqrt{{\cal K}}}\sum_n 
\frac{(\eta/\beta^2)^n}{\sqrt{(K-n)!(L-n)!}}|n_K\rangle|n_L\rangle\nonumber\\
{\cal M}&=&\sum_n \frac{|\eta/\beta^2|^{2n}}{(K-n)!(L-n)!}.
\end{eqnarray}   
Again, this state is no longer a two-mode squeezed state, but it is independent of $\phi$.  With the same argument as before, when $|\beta|$ is sufficiently large, $|\Phi_{KL}\rangle$ does approach a two-mode squeezed state,
\begin{eqnarray}\label{KL}
|\Phi_{KL}\rangle\approx\sqrt{1-\eta'^2}\sum_n 
\eta'^n|n_K\rangle|n_L\rangle,
\end{eqnarray}   
 with $\eta'\approx \eta \sqrt{KL}/\beta^2$. 
Since $\sqrt{KL}\approx |\beta|^2$, $|\eta'|\approx|\eta|$. Thus, one should not lose any entanglement by this procedure.
Let us consider more explicitly how much entanglement this procedure retains on average, as compared to the entanglement that the two-mode squeezed state possesses.
This average is given by
\begin{equation}
\bar{E}=\sum_{K,L} P(K,L)E(|\Phi_{KL}\rangle),
\end{equation}
with $P(K,L)$ the probability to find $K$ and $L$ photons, respectively, and $E(.)$ the standard measure of entanglement of pure states.
This should be compared with the entanglement present in the two-mode squeezed state $|\eta_{\phi}\rangle$ \cite{enk},
\begin{equation}
E=\cosh^2 r \log_2 (\cosh^2 r)- \sinh^2 r\log_2 (\sinh^2 r),
\end{equation}
with $\tanh r=\eta$.
For reasonable (i.e., experimentally achievable) values of $\eta$ between 0.1 and 0.5 the entanglement $E$ varies between 0.08 and 1.08 ebits.
Depending on the amplitude $\beta$ one will recover almost all of this entanglement on average.
Figure~1 shows that the larger $\beta$ is, the more entanglement is recovered, as expected, with $\beta\approx 10$ sufficient to retain more than 99\%.
Again, the intuition is that for large $\beta$ the photon number measurement hardly reveals any information about the original two-mode squeezed state, thus leaving its entanglement intact.
\begin{figure}\label{f1} \leavevmode
\epsfxsize=8cm \epsfbox{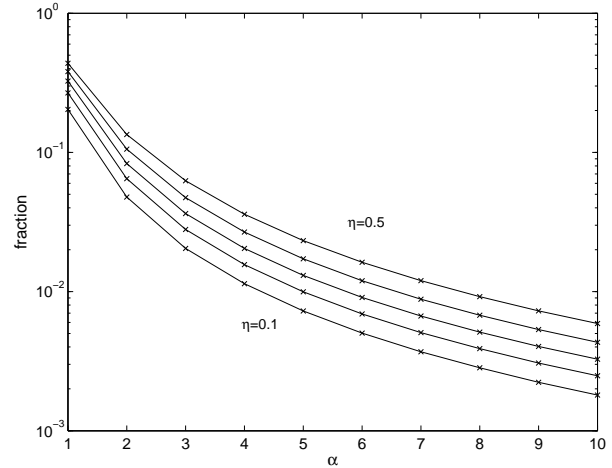} \caption{
Fraction $(E-\bar{E})/E$ of the entanglement in the state $|\eta_{\phi}\rangle$
lost on average by converting to phase-reference-free encoding
as a function of $\beta$ for $\eta=0.1,0.2,\ldots 0.5$ .
}
\end{figure}

Using the above procedures Alice and Bob could perform a
teleportation protocol with continuous variables, but without
sharing a phase reference. As a first and crucial step Alice would
produce the entangled state (\ref{ML}) and send half of it to Bob.
This is indeed what is envisaged in \cite{bartlett}, and 
appears to refute the claim
in \cite{jmo} that a shared
reference is always necessary.

However, if one now realizes that sending half of  the entangled
state $|\Phi_{KL}\rangle$ is no more efficient in terms of
resource usage than sending one mode that contained half of the
original entangled but phase-dependent state, along with a mode
that contained a coherent state with the same phase, one sees that
there is in fact not much practical difference with the standard
teleportation protocol\cite{akira}, wherein an extra laser beam is
sent from Alice to Bob to act as phase reference. In fact, if not Alice 
but Bob would do the photon number measurement, Alice really just 
sends an extra laser beam to Bob.

The differences are more
of principle: since phase references are always constructed from finite
resources, there is no way to avoid errors due to imperfection of the shared
reference. This contrasts markedly with the results (for qubits) presented
in \cite{bartlett}, where it is shown that perfect quantum protocols are in
principle possible without establishing a reference frame and with
asymptotically no resource loss.

\end{document}